\documentclass{article}

\usepackage{arxiv}

\usepackage[utf8]{inputenc} 
\usepackage[T1]{fontenc}    
\usepackage{hyperref}       
\usepackage{url}            
\usepackage{booktabs}       
\usepackage{amsfonts}       
\usepackage{nicefrac}       
\usepackage{microtype}      
\usepackage{lipsum}
\usepackage{graphicx}
\graphicspath{ {./images/} }

\usepackage[
backend=biber,
style=authoryear-icomp
]{biblatex}

\bibliography{references}
\title{Machine Learning Pipeline for Segmentation and Defect Identification from High Resolution Transmission Electron Microscopy Data}
\author{
  C.K. Groschner \\
  Department of Materials Science and Engineering\\
  University of California Berkeley\\
  Berkeley, CA 94720 \\
  \And
 Christina Choi \\
  Department of Materials Science and Engineering\\
  University of California Berkeley\\
  Berkeley, CA 94720 \\
  \And
 M.C. Scott\\
  Department of Materials Science and Engineering\\
  University of California Berkeley\\
  Berkeley, CA 94720 \\
  \\
  Molecular Foundry \\
  Lawrence Berkeley National Laboratory \\
  Berkeley, CA 94720 \\
  \texttt{mary.scott@berkeley.edu} \\
  }
\begin{document}






     




\maketitle

\begin{abstract}
In the field of transmission electron microscopy, data interpretation  often lags behind acquisition methods, as image processing methods often have to be manually tailored to individual datasets.  Machine learning offers a promising approach for fast, accurate analysis of electron microscopy data. Here, we demonstrate a flexible two step pipeline for analysis of high resolution transmission electron microscopy data, which uses a U-Net for segmentation followed by a random forest for detection of stacking faults. Our trained U-Net is able to segment nanoparticle regions from amorphous background with a Dice coefficient of 0.8 and significantly outperforms traditional image segmentation methods. Using these segmented regions, we are then able to classify whether nanoparticles contain a visible stacking fault with 86\% accuracy. We provide this adaptable pipeline as an open source tool for the community. The combined output of the segmentation network and classifier offer a way to determine statistical distributions of features of interest, such as size, shape and defect presence, enabling detection of correlations between these features. 
\end{abstract}

\keywords{HRTEM, Automated Analysis, Segmentation, Deep Learning, Structure Classification, High Throughput}



\section*{Introduction}

The need for rapid, but accurate, image analysis is ubiquitous in electron microscopy studies of nanomaterials. With the advent of fast, high efficiency electron detectors and automated imaging protocols (\cite{schorb_software_2019,lebeau_2020}), incorporating electron microscopy into high throughput materials design efforts (\cite{jain_commentary:_2013}) is increasingly feasible. These new and upcoming capabilities strongly motivate automated methods to extract relevant structural features, such as nanoparticle size, shape, and defect content, from high resolution transmission electron microscopy (HRTEM) data to link these features to bulk properties and study the influence of heterogeneity on bulk behavior (\cite{zhao_influence_2018,orfield_correlation_2015,rasool_measurement_2013}). In general, protocols which outperform classical image analysis and do not require time-consuming manual analysis are needed. Given recent advances in image interpretation using deep learning (\cite{NIPS2013_5207, navab_u-net:_2015}), segmentation \textit{via} convolutional neural net (CNN), along with other machine learning techniques, are promising routes toward automatic interpretation of HRTEM micrographs. Here, we demonstrate a two-step pipeline to detect and classify regions of interest in HRTEM micrographs. This tool uses a convolutional neural net to identify crystalline regions (nanoparticles) from an amorphous background in the images, and then feeds individual regions of interest into a random forest classifier to detect whether or not they contain a stacking fault (\cite{Groschner2020}).

Previous work has applied deep learning to atomic resolution images of nanomaterials from high resolution scanning transmission electron microscopy and HRTEM, but this has focused on atomic column segmentation or atomic column tracking and specific structure analysis (\cite{ziatdinov_deep_2017, madsen_deep_2018, Ziatdinov2019AtomicMF, Maksov2019, schiotz_2018, schiotz_2019, kalinin_lupini_dyck_jesse_ziatdinov_vasudevan_2019, Dan2019, Jany2020, Forster2020,Dennler2020}). Other work has focused on lower resolution TEM micrographs, where amplitude contrast is the dominant contrast mechanism (\cite{Groom2018,Frei2018,Laramy2015,Li2018,Guven2018,Roberts2019,Yeom2020}).  Less work has been done to identify isolated regions of interest in HRTEM data and classify the structures therein. Here, we aimed to first identify nanoparticle regions, isolate them, and then classify the nanoparticles according to their structure.

We have focused on conventional HRTEM as it is a dose-efficient and high frame rate imaging mode and therefore useful for fast data acquisition on a wide range of materials. However, segmentation in HRTEM is particularly challenging as the contrast between the substrate that the nanomaterial sits on and the nanomaterial itself can be very low. Because traditional image processing techniques are highly error prone for these types of images, we have implemented CNN for semantic segmentation, which is segmentation of the image on a pixel by pixel basis. Our CNN has a modified U-Net architecture (\cite{navab_u-net:_2015}) and can accurately segment a diverse population of nanoparticles despite a small number of training images. Segmentation is demonstrated on both gold and cadmium selenide nanoparticles. After segmentation, individual nanoparticle regions can be isolated and fed directly into existing python tools to extract size and shape statistics. To detect the presence of defects in nanoparticle regions, we implement a random forest classifier. We demonstrated the random forest classifier's ability to detect stacking faults in the CdSe subset of identified nanoparticles. Both the CNN and classifier demonstrate state of the art performance at their respective tasks. While this work focuses on HRTEM images of nanoparticles supported on a carbon substrate, in principle the tool can be used to detect any regions of crystallinity in HRTEM data, and the simple random forest classifier is designed to be easily retrained to detect a variety of types of defects, making this a flexible pipeline suited for a variety of image analysis tasks.

\begin{figure*}[hbt!]
    \centering
    \includegraphics[width=0.6\textwidth,scale=0.5]{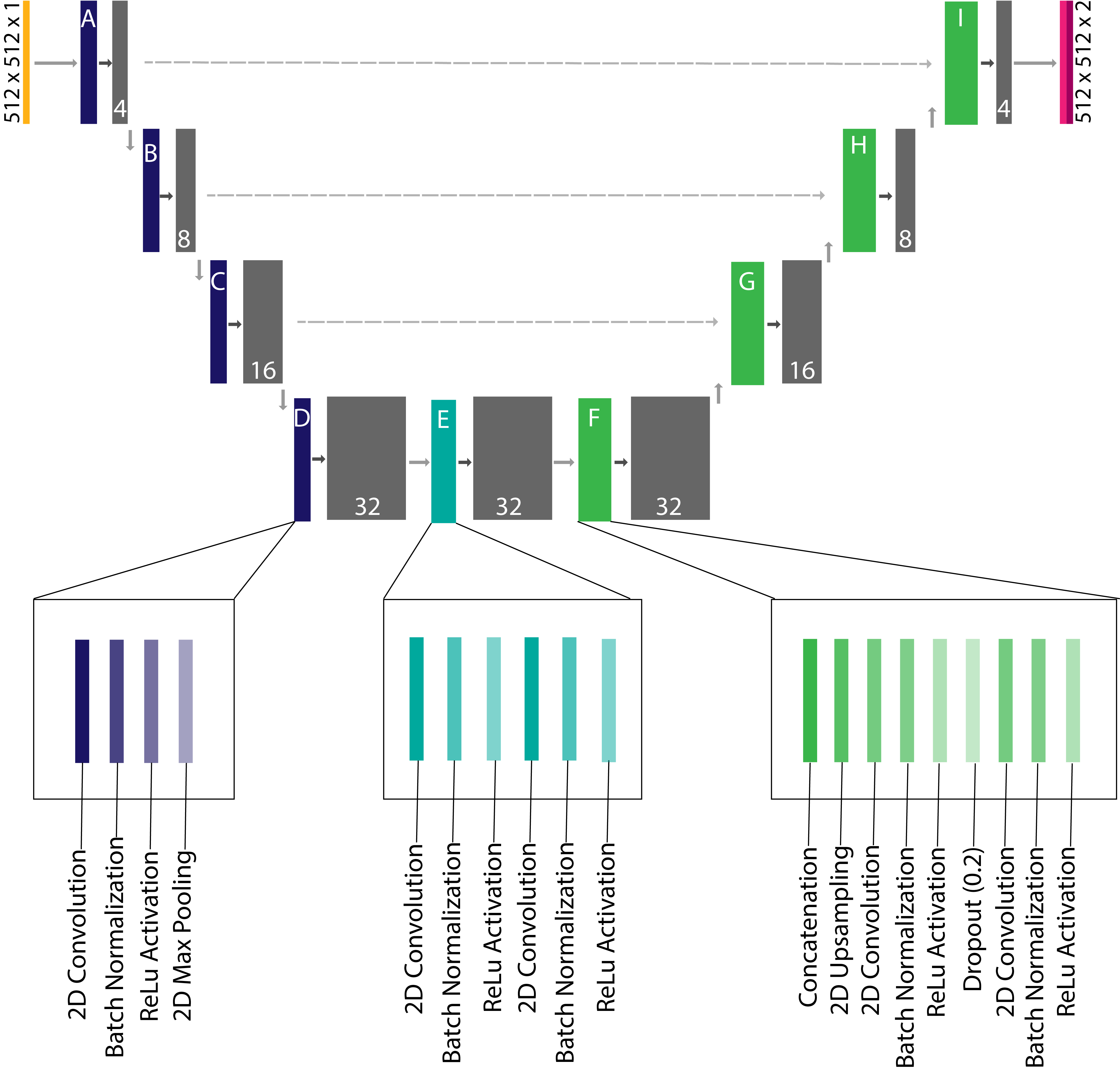}
    \caption{The U-Net style architecture of CNN implemented for segmentation of nanoparticle regions. Blocks colored gray represent the output features and are labeled with the number of features created. Residual blocks A-D contain a 2D convolutional layer, batch normalization layer, rectified linear unit (ReLu) activation layer, and a 2D max pooling layer. Residual block E contains a 2D convolutional layer, batch normalization layer, and ReLu activation layer, which is repeated once. Residual blocks F-I contain a 2D upsampling layer, concatenation layer, a 2D convolution, batch normalization, and ReLu activation followed by a dropout layer and another set of 2D convolution, batch normalization, and ReLu activation layers. The final layer is a softmax activation layer which outputs two segmentation maps one for each class, background and particle.}
    \label{fig:U-Net}
 \end{figure*}

\section*{Methods}

\subsection*{Strategy}
The overall goal of this work was to create a tool for automated structural analysis of nanoparticles from HRTEM data. In order to achieve this, we created a two-step pipeline for accurate segmentation and defect classification. The first stage of the analysis pipeline was to train a U-Net convolutional neural network to segment particle regions. The details of the architecture are shown in Figure \ref{fig:U-Net}. After segmentation, we isolated individual nanoparticle regions using a series of morphological closings and openings. These isolated regions were used for feature generation and training of a random forest classifier to detect stacking faults.

\subsection*{Data Collection and Preprocessing}
We collected 46 1024x1024 micrographs of CdSe nanoparticles and 13 4096x4096 micrographs of Au nanoparticles using an aberration corrected TEM at 300kV. To reduce the computational demands, we sliced each micrograph into four or sixteen 512x512 images. To establish a ground truth for segmentation the data was manually labeled by an experienced researcher into either a particle class or background class. The ground truth segmentation maps were created using the MATLAB labeler application (\cite{noauthor_get_nodate}). All the micrographs were then normalized by first applying a median filter with a 3x3 kernel which served to remove spurious X-rays. After filtering, the images were normalized.

 Significant portions of the raw input images were carbon background, which meant that when sliced into 512x512 images only 28\% of images contained nanoparticles.  Therefore, any image that did not contain particle class pixels was discarded from the training and test set. The remaining data was  split so that 129 images were used in the training set and 43 images were in the test set.  Images were rotated and flipped to augment the dataset to more than 1000 images which were then split 4:1 into the training and validation sets.

\subsection*{Network Architecture}
The architecture implemented was based on the U-Net architecture developed by Ronneberger \textit{et al} following the same convolution-deconvolution structure (\cite{navab_u-net:_2015}). We modified the standard U-Net structure for accurate segmentation and to prevent overfitting.

\subsection*{Network Training}

Training was performed using the Savio GPU cluster at U.C. Berkeley using a compute node with two Intel Xeon E5-2623 v3 CPU cores and an NVIDIA K80 GPU. The node had 64 GB of available RAM. The model was built using Keras with a Tensorflow backend (\cite{Keras}). Training used 129 sample images, which then fed into a Keras image augmentation generator (\cite{Keras}), which randomly rotated and flipped the images.  Each epoch, or complete pass through the training set, contained 1,000 samples created from rotating the original processed 129 micrographs and split between the training and validation sets in a 4:1 ratio.  Training was stopped once the  loss on the validation set did not decrease by 0.001 after two epochs. The model was limited to train for a maximum of ten epochs. The model used categorical cross-entropy for the loss function and Adam as the optimizer with a learning rate of 1x\(10^{-4}\) (\cite{Goodfellow-et-al-2016, Kingma2015}). Batch size was set to 20, where batch size defines how many samples from the training set are propagated through the network before updating the model weights.

\subsection*{Testing}
A holdout test dataset of 43 images was reserved for final testing after model training had completed. The predicted particle segmentation map was thresholded to 0.5 prior to metric computation because this represented the best tradeoff between precision and recall. Precision and recall versus threshold curves for the network both on Au and CdSe data are provided in the Supplementary Information. Reported metrics are the results from these test sets.

\subsection*{Alternate Methods for Segmentation}
Two standard segmentation methods were applied to the dataset to compare to the U-Net segmentation. Standard thresholding was applied using Otsu's method (\cite{Otsu1979}) to threshold out the nanoparticles based on intensity. Fourier filtering was also used to segment out crystalline regions. For this filtering, the Fourier transform of the image was multiplied by an annulus to eliminate frequencies related to the material's lattice structure. The inverse Fourier transform of the filtered image was then smoothed and thresholded to create a segmentation map. 

\subsection*{Random Forest Classifier}
After applying the neural network, the resulting segmentation masks were used to isolate particle regions. To limit the number of  agglomerated regions input into the random forest, we implemented morphological filtering on predicted segmentation maps using the scikit-image python package (\cite{scikit-image}).  Regions which were too large to be independent particles were discarded before stacking fault detection. The random forest was developed using the scikit-learn python package (\cite{scikit-learn}). 329 particle regions were in the training set and 163 regions were in the test and validation set. The training set had 71 examples of stacking faults, 156 examples of no stacking fault, 80 examples particle not on the zone axis, 60 examples of agglomerations, and 43 empty regions. The training and validation set were manually labeled. The random forest had 500 decision trees, the criterion for tree splitting was gini impurity, which calculates the probability of classifying the data incorrectly. The feature set consisted of the mean, standard deviation, and center of mass of the radially integrated Fourier transform of the particle region and the mean and standard deviation of the nanoparticle image in real space. This feature set was selected such that validation accuracy and true positive rate for each class were highest. These results lead to a small and informationally dense feature set based on summary statistics.

\section*{Results and Discussion}

\subsection*{Evaluation Metrics for Segmentation}

To evaluate the neural network’s ability to segment nanoparticles from the micrograph we used three evaluation metrics: (1) Dice coefficient (DICE), also known as F1-score, (2) precision, and (3) recall. We look at three metrics because each offers a difference weighting of error types. Dice coefficient is a standard metric for segmentation tasks and measures the union of positive pixels between the true and predicted segmentation masks normalized by the total number of positive pixels in the two groups. The Dice coefficient can be a very strict metric for small particle regions because of the weighting of true positive pixels, shown in Equation \ref{eq:1}, where TP stands for true positives, FP for false positives and FN for false negatives (\cite{Taha2015}). Fewer positive pixels overall means small errors can have a strong effect on the metric. Therefore it is also important to consider other metrics, such as precision and recall. Precision essentially measures how successful the neural network is at only selecting true particles. Recall, on the other hand, provides information on how well the network can find particles.

\begin{equation} \label{eq:1} 
    DICE = \frac{2TP}{2TP + FP +FN}
\end{equation}

\subsection*{Performance of U-Net Segmentation and Comparison with Standard Image Processing Methods}

\begin{figure*}[hbt!]
    \centering
    \includegraphics[width=0.7\textwidth,scale=1]{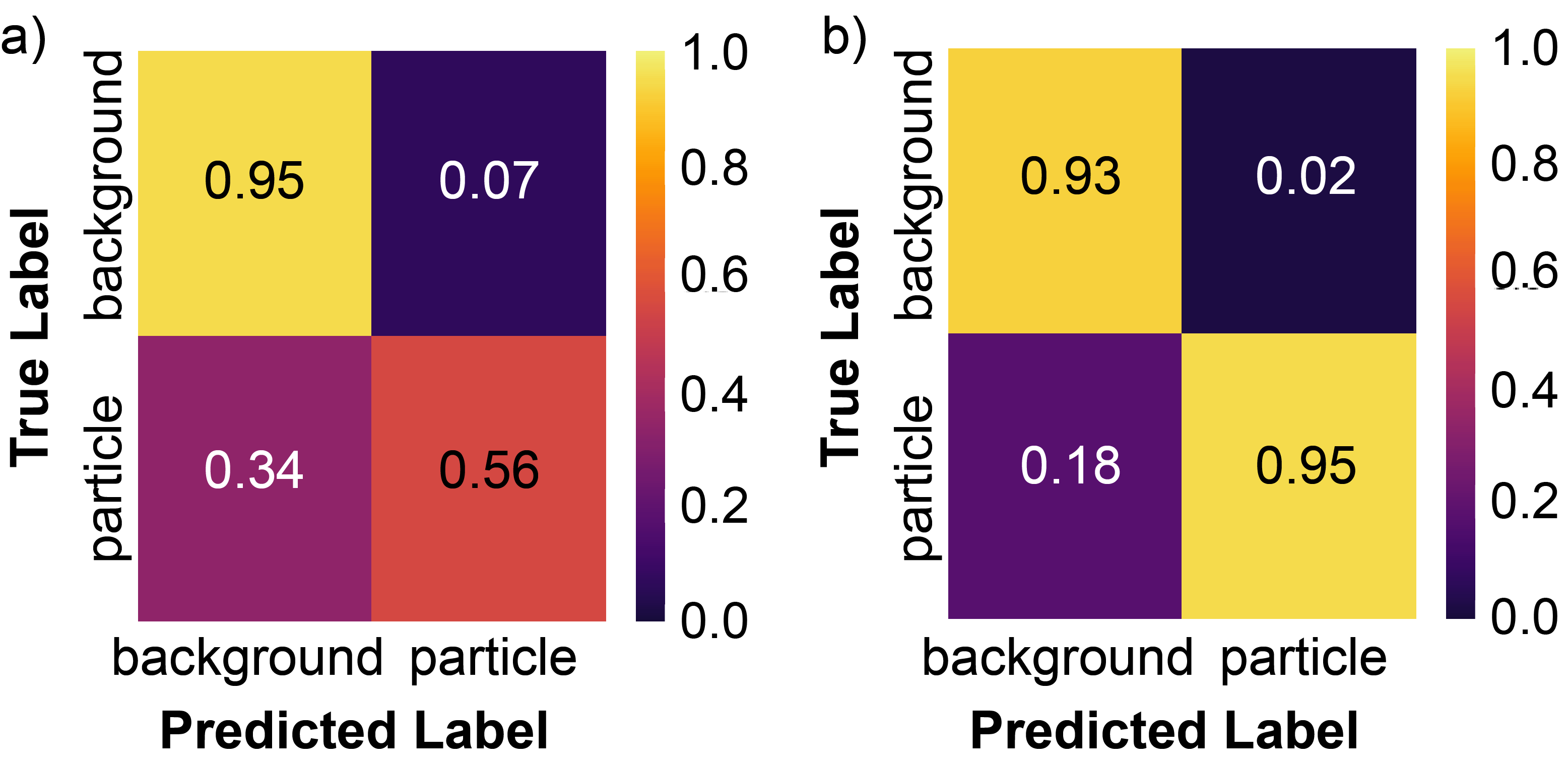}
    \caption{The confusion matrices for the trained U-Net on (a) CdSe test set and (b) for the Au test set.}
    \label{fig: cfm}
 \end{figure*}

\begin{figure*}[hbt!]
    \centering
    \includegraphics[width=0.8\textwidth,scale=1]{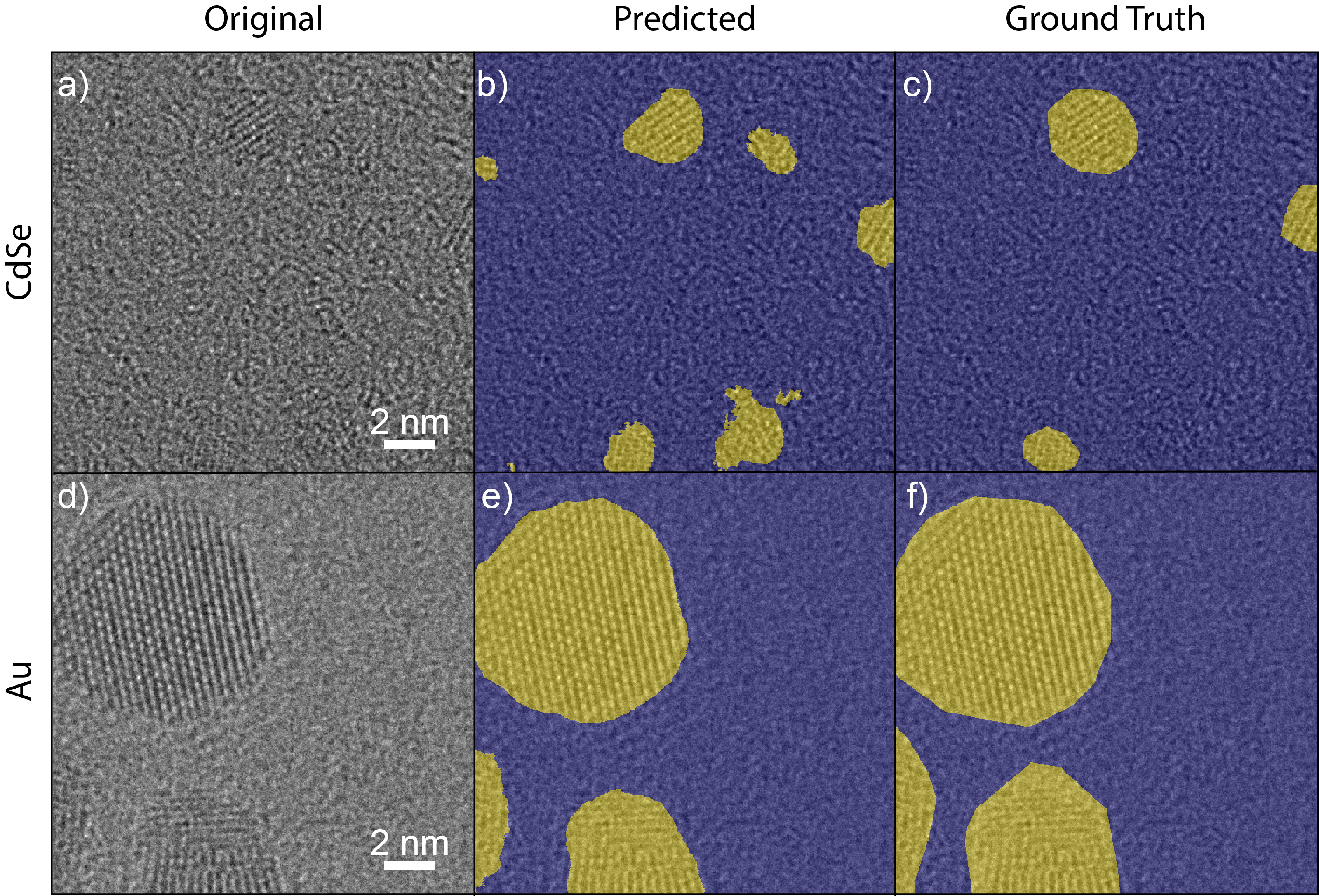}
    \caption{The top row, a-c, of the figure shows a sample micrograph, segmentation from the network and ground truth segmentation from the CdSe dataset. The second row, d-f, of the figure shows a sample micrograph, segmentation from the network and ground truth segmentation from the Au dataset. For all segmentation maps, yellow represents regions predicted to be in the particle class, blue regions predicted as background.}
    \label{fig: segmentation_example}
 \end{figure*}
We find segmentation via U-Net style CNN to be the most effective way to segment micrographs as compared to standard methods. Overall, the Dice coefficient across the test set was 0.8. Separating the test set according to sample material, the network achieved a Dice coefficient for Au nanoparticle micrographs of 0.89 while the CdSe nanoparticles were segmented with a Dice coefficient of 0.59, which is on par with other state of the art segmentation procedures (\cite{holm_2019}). Further metrics are provided in Table \ref{table:1} and the confusion matrices presented in Figure \ref{fig: cfm}. Sample segmentations for both are shown in Figure \ref{fig: segmentation_example}. CdSe particles clearly are the much more difficult segmentation case, both for human and computer labelers. CNN based segmentation is likely limited due to the limited signal relative to background and the size of the particle regions (limiting the context for the neural network to learn). We explored whether training solely on CdSe would improve the segmentation but we found that training on a diverse set of materials actually improved results. These results are presented in the Supplementary Materials. Despite a lower Dice coefficient on the CdSe data, the network created by the demonstrated training procedure results in a network which tends to under predict particle regions, as can be seen in Figure \ref{fig: cfm}(a), from the low false positive rate for the particle class. This bias is preferred to over prediction of particles which would make later classification challenging.    
 
\begin{table}[hbt!]
\begin{center} 
\begin{tabular}{|l|l|l|l|}
\hline
   & DICE & Precision & Recall \\ 
\hline
Combined Dataset & 0.8 & 0.82 & 0.78 \\ 
\hline
CdSe Data & 0.59 & 0.56 & 0.62 \\ 
\hline
Au Data & 0.89 & 0.95 & 0.84 \\
\hline
\end{tabular}
\end{center}
\caption{Performance metrics for test sets}
\label{table:1}
\end{table}

\begin{figure*}[hbt!]
    \centering
    \includegraphics[width=\textwidth,scale =1]{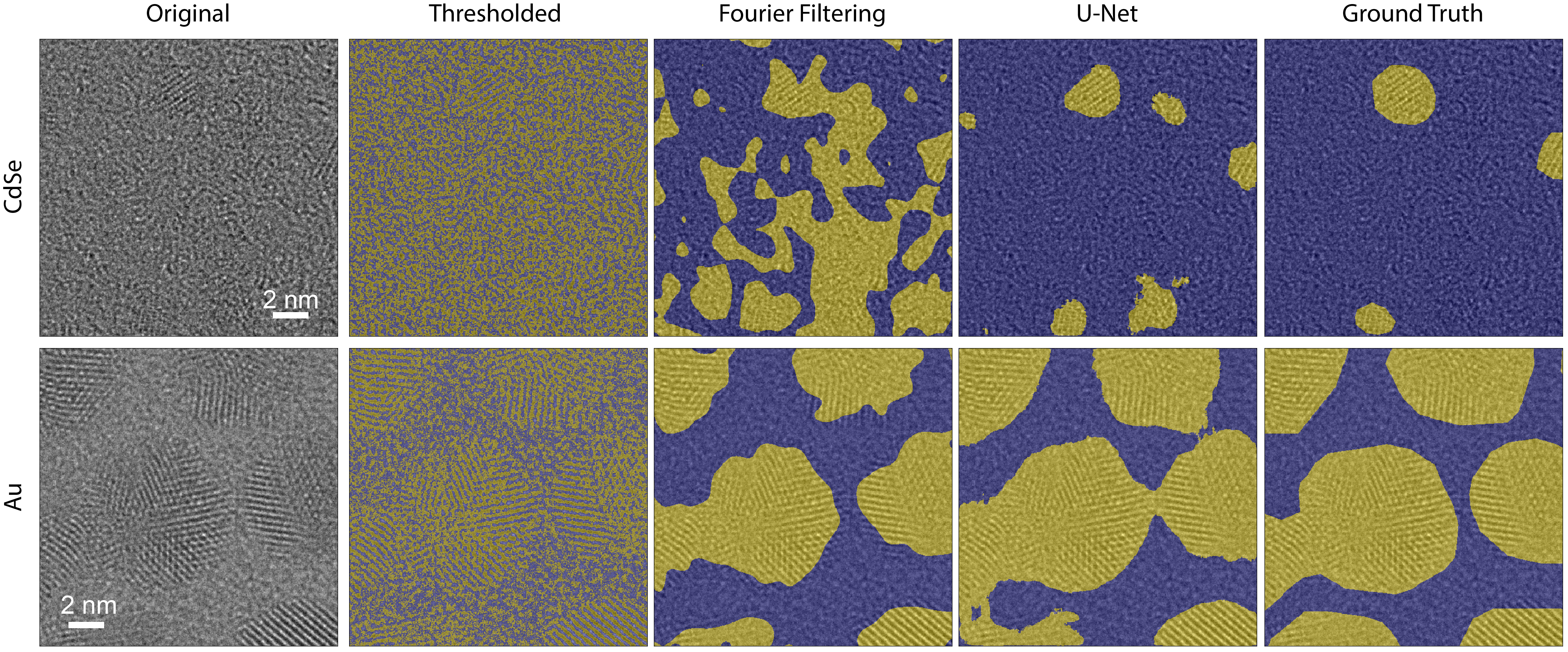}
    \caption{Sample micrographs of CdSe and Au particles and the resulting segmentation maps when segmented by Otsu's method, Fourier filtering, and U-Net. For all segmentation maps, yellow represents regions predicted to be in the particle class, blue regions predicted as background.}
    \label{fig: seg_methods}
\end{figure*}

In order to determine if the network was more successful than standard image processing techniques, we segmented the micrographs using Otsu's method (\cite{Otsu1979}) and using Fourier filtering (\cite{Buseck1988,DeJong1989}). More information for each segmentation method is supplied in the Methods section. We found that basic thresholding on the complete CdSe test set achieved a Dice coefficient of 0.21 and Fourier filtering achieved a Dice coefficient of 0.33. We note that this is significantly worse than using Otsu’s method to threshold the Au test set, which leads to a Dice coefficient of 0.45 or using Fourier filtering which leads to a Dice coefficient 0.78. The Dice coefficients for the combined test set was 0.33 and 0.52 for thresholding via Otsu’s method and Fourier filtering respectively. This takes into account optimizing the radius of the annulus mask in Fourier filtering which impacts the segmentation quality. Sample segmentations for each method are shown in Figure \ref{fig: seg_methods}. Clearly, on all accounts, the neural network performs significantly better than standard segmentation methods, but it is a particular improvement for the CdSe dataset, where there is little amplitude contrast. In addition, beyond the improved segmentation, using a neural network also has the benefit of not hand tuning parameters for each image, which are critical in standard processing methods. Therefore, for this  type of TEM data, U-Net provides a better, and more high throughput, means of particle segmentation.

\begin{figure*}[hbt!]
    \centering
    \includegraphics[width=0.9\textwidth,scale=0.5]{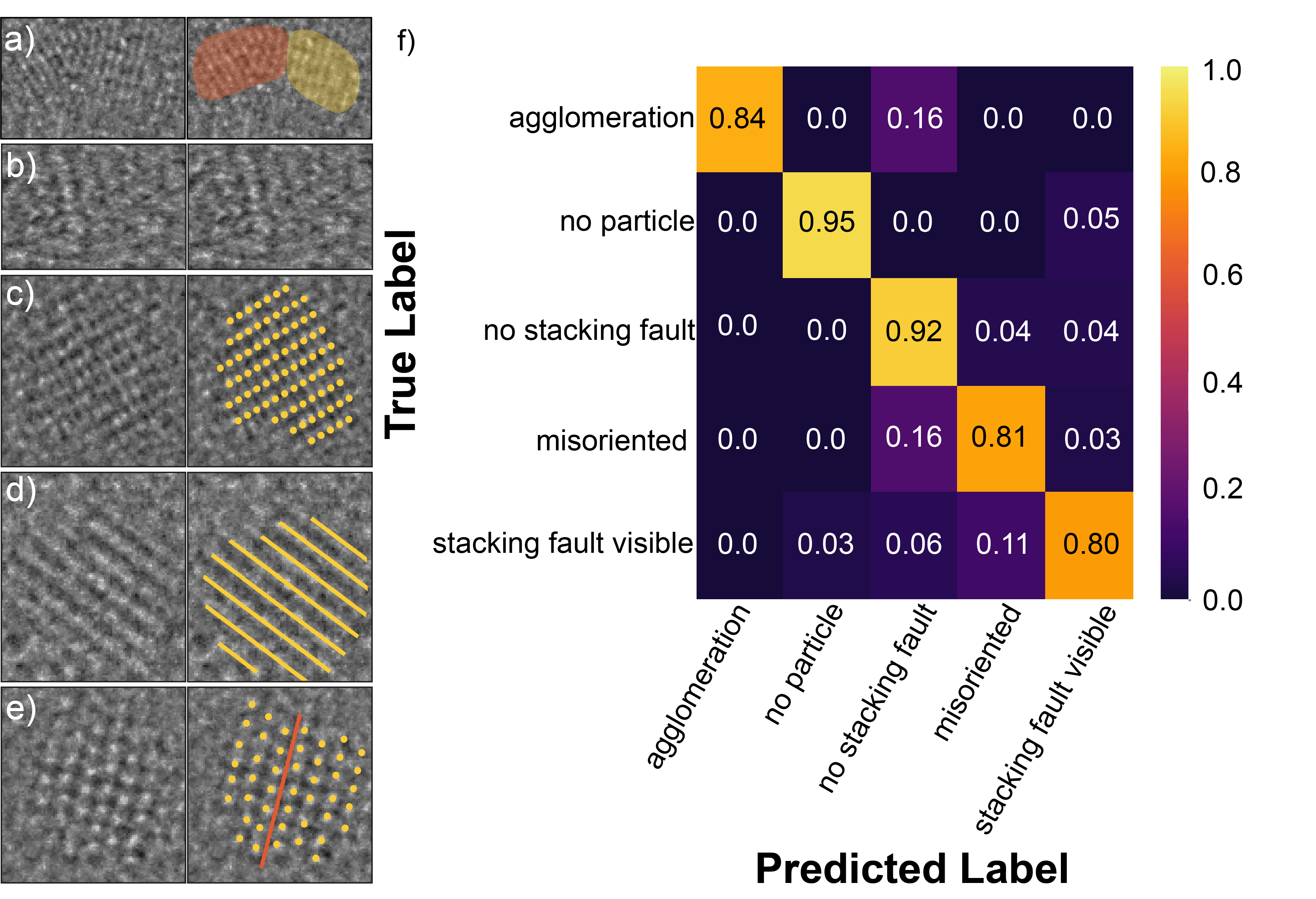}
    \caption{Sample micrographs, original on the left and annotated on the right, of the five classes predicted: a) agglomeration (the two particle regions are highlighted in orange and yellow respectively), b) no particle, c) no stacking fault (atomic columns, marked in yellow, are aligned), d) misoriented (atomic planes, marked by yellow lines, are visible but atomic columns are not), and e) stacking fault visible (atomic columns, marked in yellow, can be seen to be offset across the stacking fault, marked in orange). f) Confusion matrix for the random forest classifier.}
    \label{fig: rf_cfm}
 \end{figure*}

\subsection*{CdSe Particle Statistics from Random Forest Classification}
The primary reason to develop a highly accurate segmentation technique for HRTEM is to enable high throughput analysis of crystal structures. Segmented regions of interest can be used in further classification models and automated statistics. Statistics on size and shape can be provided directly from segmentation results. Segmented regions can also be passed to classifiers for structure or defect detection as we will demonstrate.
 
Stacking faults in CdSe are of interest due to the influence they can have on particle growth and therefore shape of CdSe nanoparticles (\cite{Peng2000}). Here we show that a simple random forest can be used to detect stacking faults. A random forest was chosen for several reasons, including the excellent performance of ensemble methods as well as the  speed and simplicity to train (\cite{geron_aurelien_hands-machine_2017}). Combined with high model interpretability, makes this classifier a good choice for fast iteration and rapid development. 

The random forest demonstrated is used to categorize the CdSe nanoparticles into five different classes: (1) particle has a stacking fault visible, (2) particle does not have stacking fault visible, (3) particle does not have atomic column contrast (misoriented), (4) region is an agglomeration of particles, and (5) region is background. Sample micrographs for each class are shown in Figure \ref{fig: rf_cfm}(a-e). We included the last category to handle the few cases in which the neural network segments  a region incorrectly, since perfect segmentation cannot be guaranteed from our network.  The feature set we developed for the random forest is extremely simple and consists only of the mean, standard deviation, and center of mass of the radially integrated Fourier transform of the particle region and the mean and standard deviation of the nanoparticle image in real space. Figure \ref{fig: rf_cfm}(f) shows the confusion matrix for the random forest classifier on the test data set. The confusion matrix provides the fraction of correct and incorrect labels for each class. The random forest achieves 86\% class balanced accuracy compared to the expert labels given. The random forest predicted that 52\% of particles were in the correct orientation to observe a stacking fault, and of those 38\% of particles contain a stacking fault. This can be compared to the ground truth of 49\% of particles which were in the correct orientation to observe a stacking fault and the 35\% of those particles that contained a stacking fault. As a consequence, we can see that using an extremely limited feature space still yields good classification of particles. 

In summary, this two step pipeline offers a flexible method to identify crystalline regions of interest and classify individual regions according to known features. The segmentation portion of the pipeline outperforms standard segmentation methods, even when those methods are optimized for individual datasets. The easily retrainable random forest classifier enables detection of known features within the identified regions. We note that the outputted individual regions could also be input into unsurpervised classifiers for detection of previously unidentified features. Further work could also aim to implement more complex structure and defect classification methods(\cite{li2018automated, Maksov2019}). One particularly useful future area of work would be to develop zone axis classification alongside defect identification, which would aid in the interpretation of observed defects. 

Coupled with existing size and shape analysis tools, the output of the segmentation network and classifier demonstrated here offer a way to determine statistical distributions of features of interest, such as size, shape and defect presence, and importantly, allow one to detect correlations between these features. This capability will be critical to offer insight into nanomaterial population evolution during  during high throughtput synthesis studies, or to identify key structural features that influence bulk properties in combinatorial synthesis efforts.

\section*{Conclusion}

Here we have demonstrated a  method of flexible automated analysis for nanoparticles in HRTEM by combining a neural network for segmentation and a random forest for classification tasks.  We have shown that a U-Net architecture can far outperform traditional image processing techniques for segmentation of HRTEM images, while a simple classification tool can accurately classify structural features using a very limited number of parameters.  Breaking apart the segmentation and classification tasks leads to an accurate tool with limited data labelling and feature engineering requirements. Moreover, this pipeline provides a flexible open source tool as the base for further analysis and classification tools for local atomic structure. Such quantitative, automated analysis will have significant implications for a broad range of nanoparticle synthesis, structure-property relationship, and combinatorial synthesis studies. 

\section*{Acknowledgements}

Work at the Molecular Foundry was supported by the Office of Science, Office of Basic Energy Sciences, of the U.S. Department of Energy under Contract No. DE-AC02-05CH11231. This material is based upon work supported by the National Science Foundation Graduate Research Fellowship under Grant No. DGE-1752814. This work was also supported by National Science Foundation STROBE grant DMR-1548924. CdSe synthesis was supported by the Department of Energy, Office of Energy Efficiency and Renewable Energy (EERE), under Award Number DE-EE0007628. The authors would like to thank Haoran Yang and Emory Chan for providing the CdSe nanoparticles. The authors would also like to thank Colin Ophus, Mike MacNeil, and Tess Smidt for their feedback and comments during the writing of this manuscript.

\section*{Author contributions statement}

M.C.S. and C.K.G. conceived the project and designed the experiments. C.K.G. and C.C. optimized the network architecture. C.K.G. acquired the training images, trained the network, and tested the network. C.K.G. and M.C.S. wrote the manuscript. All authors discussed the results and commented on the manuscript.

\section*{Additional information}

\subsection*{Data Availability}

The complete workflow for micrograph preprocessing, network training, and testing is available in Jupyter notebooks at
\url{https://github.com/ScottLabUCB/HTTEM/tree/master/pyNanoFind}. All micrographs and ground truth segmentation maps are available on Zenodo: \url{https://doi.org/10.5281/zenodo.3755011}.

\subsection*{Competing Interests}
The authors declare no competing financial or non-financial interests.

\printbibliography
\end{document}